# Superconductivity dichotomy in K-coated single and double unit cell FeSe films on SrTiO$_3$

Chenjia Tang,[1] Ding Zhang,[1] Yunyi Zang,[1] Chong Liu,[1] Guanyu Zhou,[1] Zheng Li,[1] Cheng Zheng,[1] Xiaopeng Hu,[1] Canli Song,[1,2] Shuaihua Ji,[1,2] Ke He,[1,2] Xi Chen,[1,2] Lili Wang,[1,2*] Xucun Ma[1,2] and Qi-Kun Xue[1,2†]

[1] *State Key Laboratory of Low-Dimensional Quantum Physics, Department of Physics, Tsinghua University, Beijing 100084, China*

[2] *Collaborative Innovation Center of Quantum Matter, Beijing 100084, China*

We report the superconductivity evolution of one unit cell (1-UC) and 2-UC FeSe films on SrTiO$_3$(001) substrates with potassium (K) adsorption. By *in situ* scanning tunneling spectroscopy measurement, we find that the superconductivity in 1-UC FeSe films is continuously suppressed with increasing K coverage, whereas non-superconducting 2-UC FeSe films become superconducting with a gap of ~17 meV or ~11 meV depending on whether the underlying 1-UC films are superconducting or not. This work explicitly reveals that the interface electron-phonon coupling is strongly related to the charge transfer at FeSe/STO interface and plays vital role in enhancing Cooper pairing in both 1-UC and 2-UC FeSe films.

The highest superconducting transition temperature ($T_C$) for bulk iron-based high temperature superconductors is 56 K to date [1]. Single unit cell (1-UC) thick FeSe films grown on SrTiO$_3$(STO)(001) surface with a superconducting gap Δ ~20 meV [2] exhibit not only possible higher $T_C$, as evidenced by both *ex-situ* and *in-situ* transport measurements [3-7], but also very peculiar electronic structure. The Fermi surface of the superconducting 1-UC FeSe film on STO is simple and consists only of electron pockets at the Brillouin zone corners, compared to coexisting electron and hole pockets for extensively annealed 2-UC and thicker FeSe films on STO as well as bulk iron-based superconductors [8-11]. As a result, the 2-UC and thicker films on STO do not superconduct at temperature down to 4 K [2]. The absence of hole pockets on the Fermi surface of 1-UC FeSe films indicates heavy electron doping [8-11], which is believed to originate from the oxygen vacancies in STO substrates and lift the Fermi level significantly in the bulk band structure [10, 12]. The presence of hole pockets in multilayer FeSe films therefore reflects an insufficient doping from STO. Indeed, recent ARPES [13, 14] and scanning tunneling spectroscopy (STS) [15] studies demonstrate that the superconductivity in multilayer FeSe films on STO could be realized by electron doping with potassium (K) adsorption. Superconducting gaps Δ ~10 meV which close at 44 ± 2 K were observed for K-coated 4-50 UC films [14]. On the other hand, the observations that the magnitude of the gap increases monotonically with decreasing thickness from 4-UC down to 2-UC film and the largest gap is achieved in 1-UC FeSe [2], indicate that the STO substrates have played other important role in enhancing the Cooper pairing, in addition to providing electrons. Enhanced electron-phonon coupling by FeSe/STO interface has been proposed [11, 15] and supported by several theoretical studies [16-18].

In order to address the respective roles of electron doping and FeSe/STO interface, in this work we conduct spectroscopy measurements of both superconducting and non-superconducting 1-UC and 2-UC FeSe films on STO and investigate their response to K adsorption. It is found that K adsorption always suppresses the superconductivity in 1-UC FeSe films. On the other hand, K adsorption can switch on superconductivity in 2-UC FeSe films no matter the underlying 1-UC FeSe films are superconducting or not. The gap size in the former case (~17 meV) is evidently larger than that in the latter case (~11 meV). Although in both cases superconductivity is realized after electron doping, the results explicitly reveal the important role of the FeSe/STO interface. This interface enhanced effect is further supported by observation of a superconducting gap of 14.5 meV in half UC $K_xFe_2Se_2$ films grown on 1-UC FeSe/STO, which is nearly twice of the gap value of the corresponding bulk materials [19].

The FeSe thin films were grown on $TiO_2$ terminated Nb-doped STO(001) substrates by molecular beam epitaxy (MBE), using the method described in our previous work [2]. To directly compare the properties of 1-UC and 2-UC films, we prepared FeSe films with a nominal thickness of 1.3 UC to obtain 1-UC and 2-UC thick films simultaneously. The as-grown films were not superconducting due to the Se-rich growth condition and superconductivity in 1-UC FeSe films could only be achieved after extensive annealing [9, 20, 21]. Here, we controlled the property of the films by gradually increasing the annealing temperature from 420 °C, 430 °C, 450 °C to 470 °C. Correspondingly, 1-UC FeSe films are non-superconducting (bottom blue curve in Fig. 1(c)) at 420 °C, and superconducting with an initial gap $\Delta_{ini}$ ~12 meV at 430 °C (not shown), $\Delta_{ini}$ ~13 meV at 450 °C (bottom blue curve in Fig. 1(d)) and $\Delta_{ini}$ ~17 meV at 470 °C (bottom curve in Fig. 1(f)). 2-UC FeSe films remain non-superconducting (lower red curves in Fig. 1(c) and 1(d)), consistent with previous results [2, 10, 11]. We then deposited K atoms on the samples as previously reported [15] and performed *in-situ* STM/STS measurements at 4.6 K and ARPES measurement at 70 K to investigate electronic and superconductivity properties upon K adsorption.

Figures 1(a) and 1(b) show the morphology of K-coated FeSe films. Potassium atoms adsorb randomly on the surface of 1-UC and 2-UC FeSe films (Fig. 1(a)), with some local 2 × 2 and √5×√5 reconstructions at coverage of 0.16 ML (Fig. 1(b)). The two reconstructions correspond to K coverage of 0.25 ML and 0.20 ML, respectively. This is consistent with the previous observation that K atoms adsorb individually on the surface below a coverage of ~0.20 ML and form clusters above this coverage [15]. Once clusters are formed, further adsorption contributes little electrons. Hence, 0.20-0.25 ML is the optimal doping coverage [15]. Here, 1 ML is defined as the coverage at which K atoms occupy all the hollow sites of Se lattice to form stoichiometric $K_1Fe_2Se_2$ [22].

When the 1-UC FeSe films are non-superconducting, K-adsorption fails to induce superconductivity, as shown by the blue curves in Fig. 1(c). For superconducting 1-UC FeSe films, K adsorption always suppresses superconductivity. The situation is illustrated using two different samples that were annealed at 450°C (Fig. 1(d)) and 470°C (Fig. 1(f)), respectively. The sample annealed at 450 °C exhibits an initial gap $\Delta_{ini}$ ~13 meV (lower blue curve in Fig. 1(d)), 0.1 ML K nearly kills its superconductivity as coherence peaks almost vanish (upper blue curve in Fig. 1(d)). The sample annealed at 470 °C has larger gap ($\Delta_{ini}$ ~17 meV), which systematically decreases with increasing K adsorption and becomes ~11 meV at 0.15 ML (Fig. 1(f)). At 0.17 ML, the coherence peaks disappear and no well-defined gap can be observed. Contrast to the continuous suppression of superconductivity, the extent of electron doping increases monotonically with K adsorption, as seen by a continuous downward shift of the kink in valence band (Fig. 1(e)).

The electron doping with K adsorption is further identified by ARPES study. We measured a sample with mixed normal and superconducting phases (annealed at 450 °C) and named the bands according to previous ARPES study [9]. The mixed states are characterized by a weak hole band N1 as well as a strong hole band S1 near Γ point (Fig. 2(a)) and an electron band S2 near M point (2(c)). After 0.20 ML K adsorption, all bands shift downwards significantly, indicating electron doping upon K adsorption (Fig. 2(b) and 2(d)). The amount of doped electrons estimated from the electron-like Fermi surface size around M point is 0.04 electrons per Fe atom.

The continuous suppression of the superconductivity in 1-UC FeSe films which is independent of the initial state is superconducting or not speaks against the doping induced dome-like phase diagram that was observed in multilayer FeSe films on STO [14]. The uniqueness of single UC films on STO is further characterized by the fact that neither band splitting signaling the nematic order nor band flattening is observed (Fig. 2). Hence, the doping dependent correlation strength that determines the electronic properties of multilayer films as stated in Ref. 14 seems unrelated in 1-UC films. We further note that the superconductivity was enhanced when electrons were injected into the FeSe layer by applying field effect with STO as a back gate [20]. Following the scenario that the coupling between FeSe electrons and polar phonon from STO substrate is boosted by the charge transfer from STO to FeSe and the formation of interface electric dipole [11, 17, 18, 23], the present results indicate that the electrons doped with K adsorption from the opposite side of the FeSe layer may counteract the electric dipoles and thus weaken the Cooper pairing strength. Moreover, the adsorbed K atoms induce disorder on the surface where STS was conducted, which can also suppress the superconductivity [15].

For 2-UC films on STO, in addition to switching on the superconductivity by K-adsorption as reported previously[15], we further unveil that the magnitude of their superconducting gap depends on the property of the underlying 1-UC FeSe. When the 1-UC FeSe is initially not superconducting (lower blue curve in Fig. 1(c)), 0.1 ML K induces a superconducting gap $\Delta$ ~11 meV in the 2-UC FeSe (upper red curve in Fig. 1(c)). In the case that the 1-UC FeSe is superconducting, the K-coated 2-UC can exhibit a superconducting gap as large as the initial gap of the underlying 1-UC FeSe. As shown in Fig. 1(g), when the 1-UC films initially exhibit gaps of 14.0 ± 1.0 meV, the 2-UC films at optimal doping host gaps of 14.6 ± 0.7 meV. This is also evidenced from a similar maximum gap of ~17 meV (inset of Fig. 1(g)). The gap evolution of a 2-UC FeSe films annealed at 430 °C is also presented in Fig. 1(g). It shows similar dome-like phase diagram while the gap at optimal doping is ~12 meV, equal to the gap of 1-UC films annealed at 430 °C again. Assuming the gap of 11 meV observed when the underlying 1-UC films are non-superconducting is solely induced by electron doping, the larger gap of the 2-UC FeSe on the superconducting 1-UC films explicitly indicates that the additional role of FeSe/STO interface extends to 2-UC as well. The corresponding enhancement in the gap is 55% (from 11 mev to 17 meV), consistent with the estimation from ARPES study [11].

Finally, we show that the FeSe/STO interface could also enhance the superconductivity in $K_xFe_2Se_2$ films. After annealing at ~400 °C of the sample with $\Delta_{ini}$ ~17 meV and 0.2 ML of K, the second UC FeSe film change to half UC $K_xFe_2Se_2$, as evidenced by a step height of 0.7 nm (Fig. 3(a)) and the characteristic $\sqrt{2}\times\sqrt{2}$ reconstruction (Fig. 3(b)). Intriguingly, a spatially uniform superconducting gap of 14.5 meV is observed on the $K_xFe_2Se_2$ films, which is significantly larger than $\Delta$ ~7 meV of bulk $K_xFe_2Se_2$ [19], $\Delta$ ~4 meV for $K_xFe_2Se_2$ films on graphene [24], and $\Delta$ ~9 meV for thicker $K_xFe_2Se_2$ films on STO [25]. Given that bulk $K_xFe_2Se_2$ is heavily electron doped, the enhancement observed here should be mainly due to interface effect.

In summary, we observed a dichotomy of superconductivity in 1-UC and 2-UC FeSe films on STO upon K adsorption: the superconductivity of 1-UC films is always suppressed while the non-superconducting 2-UC films become superconducting. The superconductivity of K-coated 2-UC films depends on the initial state of the underlying 1-UC films due to the existence of the FeSe/STO interface. We also show the superconductivity of half UC $K_xFe_2Se_2$ is significantly enhanced when it is grown on FeSe/STO. The above results can be consistently understood under the interface enhanced electron-phonon coupling scenario.

* liliwang@mail.tsinghua.edu.cn, † qkxue@mail.tsinghua.edu.cn

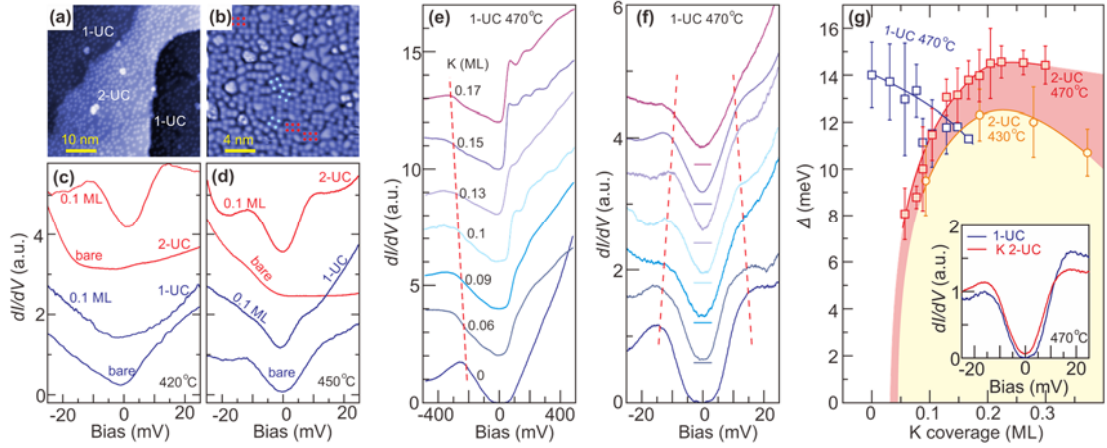

FIG. 1 (color online) Topographic images of FeSe films on STO after (a) 0.10 ML and (b) 0.16 ML K adsorption ($V = 1$V, $I = 50$ pA). The red and light blue dots in (b) illustrate the $2 \times 2$ and $\sqrt{5}\times\sqrt{5}$ reconstructions, respectively. (c) and (d) show the typical $dI/dV$ curves ($V = 30$ mV, $I = 100$ pA) taken on 1-UC (blue) and 2-UC (red) FeSe films after annealing at 420 °C and 450 °C, respectively. (e) and (f) show the typical $dI/dV$ curves ((e): ($V = 500$ mV, $I = 100$ pA) and (f): ($V = 30$ mV, $I = 100$ pA)) taken on the 1-UC FeSe films after annealing at 470°C at various K coverage. The horizontal bars in (f) indicate zero conductance position of each curve. Dashed lines are guide for the eye, marking the shift of kinks of valence band and coherence peaks. (g) The dependence of the superconducting gaps of 1-UC and 2-UC FeSe films on K coverage. Error bars are estimated from the standard deviation of $\Delta$ in the numerical fitting. The inset in (g) shows the dI/dV curves taken on 1-UC FeSe and K-coated 2-UC FeSe films under optimal doping showing nearly identical superconducting gap.

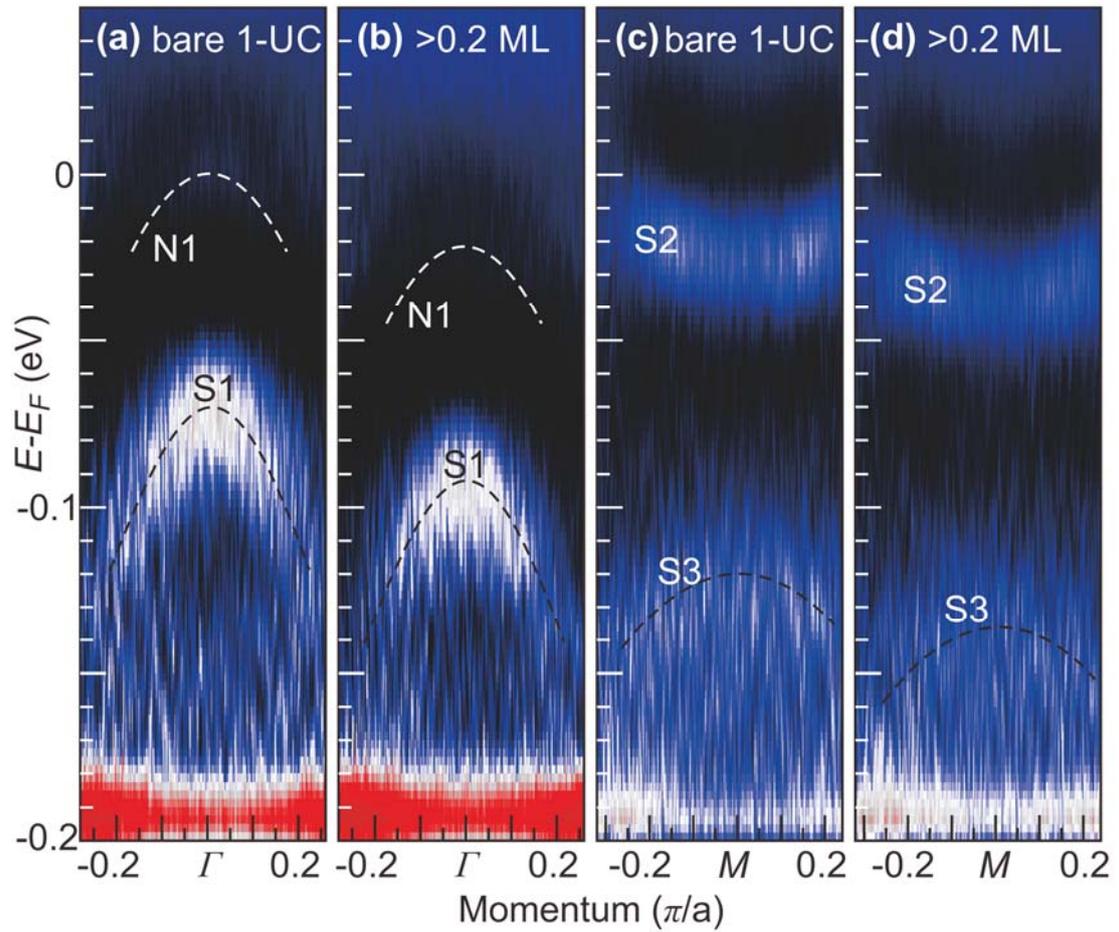

FIG. 2 (color online) Band structures near the Γ (a) and M (c) points of bare 1-UC FeSe films after annealed at 450 °C. (b) and (d) show the corresponding band structures after K adsorption at coverage larger than 0.2 ML. The images are obtained by the second derivative of the original data with respect to the energy.

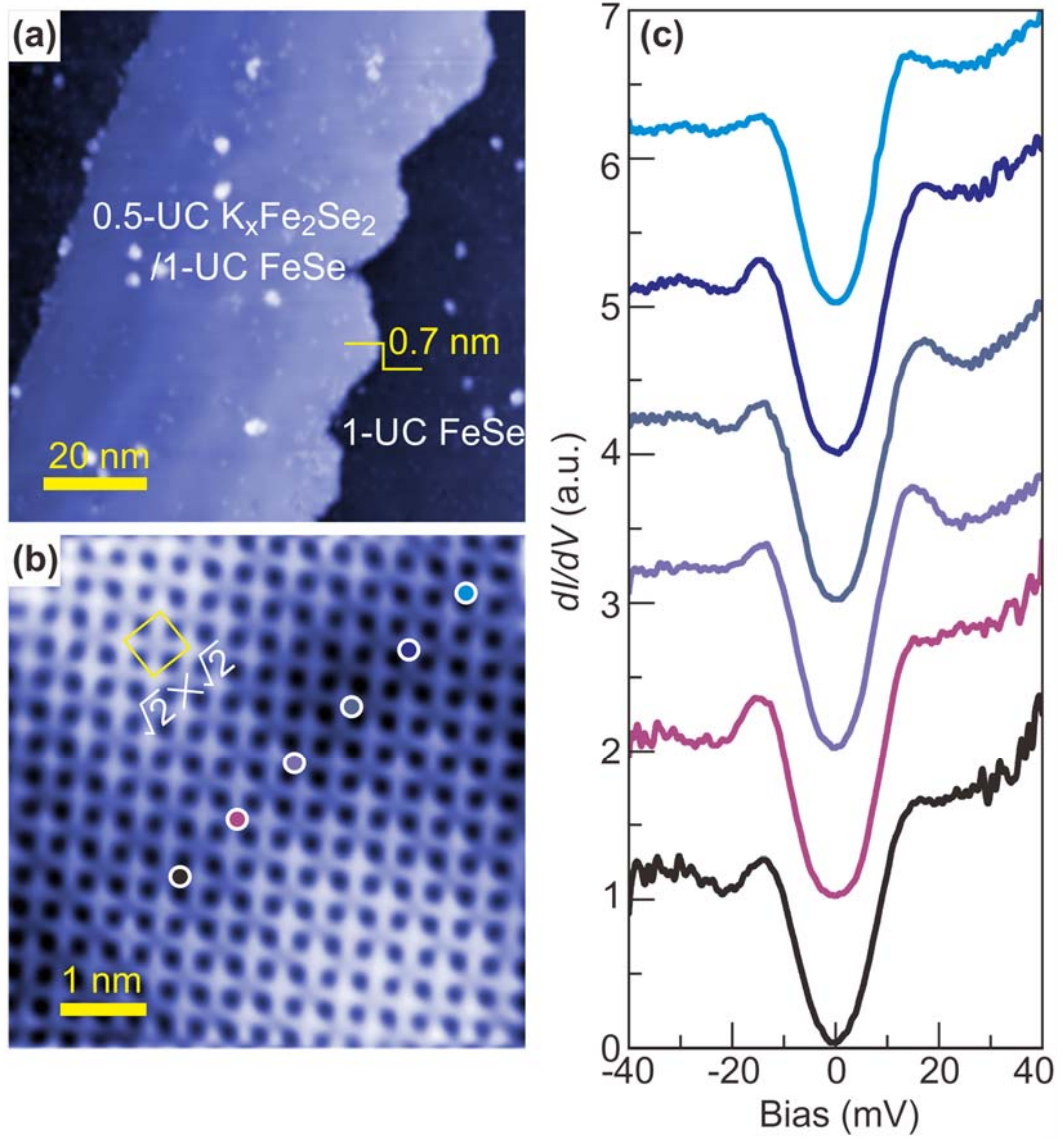

FIG. 3 (color online) (a) Topographic images ($V = 1$V, $I = 50$ pA) of the half UC $K_xFe_2Se_2$ films formed on 1-UC FeSe films. (b) Atomically resolved image ($V = 100$ mV, $I = 50$ pA) showing $\sqrt{2}\times\sqrt{2}$ reconstruction. (c) Typical $dI/dV$ curves ($V = 30$ mV, $I = 100$ pA) taken on $K_xFe_2Se_2$ films at the positions labeled by dots in (b) showing spatially uniform superconducting gap of 14.5 meV.